\documentclass[twocolumn,letterpaper]{IEEEAerospaceCLS}  

\usepackage{amsmath,amsfonts,amssymb}
\usepackage{xfrac,bm}
\usepackage[]{graphicx}    
\newcommand{\ignore}[1]{}  

\DeclareMathOperator{\sgn}{{\mathrm{sgn}}}

\newcommand{\ud}{{\mathrm{d}}}
\newcommand{\x}{{\bm{x}}}

\newcommand{\f}{{\bm{f}}}

\newcommand{\bmlambda}{{\bm{\lambda}}}

\newcommand{\mbx}{{\mathbf{x}}}

\newcommand{\mbA}{{\mathbf{A}}}
\newcommand{\mbB}{{\mathbf{B}}}
\newcommand{\mbD}{{\mathbf{D}}}
\newcommand{\mbP}{{\mathbf{P}}}

\newcommand{\mbQ}{{\mathbf{Q}}}

\newcommand{\cA}{{\mathcal{A}}}
\newcommand{\cB}{{\mathcal{B}}}

\newcommand{\cD}{{\mathcal{D}}}

\newcommand{\cH}{{\mathcal{H}}}

\newcommand{\cJ}{{\mathcal{J}}}

\newcommand{\cS}{\mathcal{{S}}}

\newcommand{\cU}{{\mathcal{U}}}

\newcommand{\cX}{{\mathcal{X}}}

\newcommand{\bR}{{\mathbb{R}}}

\newtheorem{problem}{Problem}

\newtheorem{assumption}{Assumption}

\begin{document}
\title{Feedback Strategies for Hypersonic Pursuit of a Ground Evader}

\author{%
Yoonjae Lee,~Efstathios Bakolas,~Maruthi R. Akella\\
Aerospace Engineering and Engineering Mechanics\\
The University of Texas at Austin\\
Austin, TX 78712\\
yol033@utexas.edu,~bakolas@austin.utexas.edu,~makella@mail.utexas.edu
}

\maketitle

\thispagestyle{plain}
\pagestyle{plain}

\maketitle

\thispagestyle{plain}
\pagestyle{plain}

\begin{abstract}
In this paper, we present a game-theoretic feedback terminal guidance law for an autonomous, unpowered hypersonic pursuit vehicle that seeks to intercept an evading ground target whose motion is constrained in a one-dimensional space. We formulate this problem as a pursuit-evasion game whose saddle point solution is in general difficult to compute onboard the hypersonic vehicle due to its highly nonlinear dynamics. To overcome this computational complexity, we linearize the nonlinear hypersonic dynamics around a reference trajectory and subsequently utilize feedback control design techniques from Linear Quadratic Differential Games (LQDGs). In our proposed guidance algorithm, the hypersonic vehicle computes its open-loop optimal state and input trajectories off-line and prior to the commencement of the game. These trajectories are then used to linearize the nonlinear equations of hypersonic motion. Subsequently, using this linearized system model, we formulate an auxiliary two-player zero-sum LQDG which is effective in the neighborhood of the given reference trajectory and derive its feedback saddle point strategy that allows the hypersonic vehicle to modify its trajectory online in response to the target's evasive maneuvers. We provide numerical simulations to showcase the performance of our proposed guidance law.
\end{abstract}

\tableofcontents

\section{Introduction}

Trajectory optimization for hypersonic re-entry vehicles is in general a challenging problem that does not admit analytic solutions  due to the vehicle's highly nonlinear dynamics. For this reason, most research efforts in this area have focused on developing or enhancing numerical optimal control techniques, including direct \cite{benson2006direct} and indirect \cite{vedantam2020multi} methods, for efficient trajectory optimization. In the presence of parametric uncertainty or external disturbances, however, trajectories pre-computed offline may no longer be optimal during flight, which raises the need for a hypersonic vehicle to have an inner-loop feedback guidance system for stabilization or tracking. Traditionally, neighboring optimal (or extremal) control \cite{bryson2018applied} has played an important role in aerospace applications and control of nonlinear systems which are subject to perturbations such as within initial states or terminal constraints. For the applications of neighboring optimal control in hypersonic trajectory design, one can refer to, for instance, \cite{eisler1994guidance,bain1993robust}.

Robust control and guidance problems for hypersonic vehicles have been addressed by several different methods in the literature including LQR-based feedback guidance \cite{carson2006optimal}, desensitized optimal control \cite{makkapati2021desensitized}, and deep learning method \cite{shi2021onboard} to name but a few. Most of the recent work in hypersonic guidance are, however, focused on steering a hypersonic vehicle to a target that is either static or moves in a known/deterministic fashion but is not capable of maneuvering or evading. If the target can evade, the hypersonic trajectory optimizaiton problem evolves into a differential game \cite{isaacs1965differential}, in which we are interested in finding a feedback strategy for the hypersonic vehicle to reshape its trajectory on the fly in response to the evading actions of the target.

Differential game theory, first proposed by Isaacs in his pioneering work \cite{isaacs1965differential}, has been widely used in aerospace, defense, and robotics applications as a framework to model adversarial interactions between two or more players. Pursuit-evasion games, a special class of differential games in which a group of players attempt to capture another group of players, provide a powerful tool to obtain the worst-case strategies for intercepting a maneuvering target \cite{turetsky2003missile,battistini2014differential}. Applying differential game theory to the problem of intercepting an evader by a hypersonic pursuer, however, to our best knowledge, has not yet been proposed in the relevant literature. This is due to analytical and computational complexities of the problem which can be attributed to the nonlinear dynamics of the hypersonic vehicle and the dependence of the atmospheric density upon the altitude. Although numerical methods for computing an open-loop representation of the feedback saddle point solution in two-player differential games have been well studied \cite{horie2006optimal}, such open-loop solutions can be ineffective against the target's unexpected evasion.

The main contribution of this work is a novel formulation of the problem of intercepting a maneuvering ground target by a hypersonic pursuit vehicle as a tractable pursuit-evasion game. After solving for the target's optimal evasion strategy based on a simplified version of the original nonlinear differential game, we reduce the latter game into an optimal control problem which can readily be solved with existing numerical techniques (either direct or indirect). Thereafter, we linearize the nonlinear hypersonic dynamics around the reference game trajectory and in turn formulate an auxiliary differential game whose feedback saddle point solution can be obtained analytically and computed quickly (real-time) and efficiently compared to the original noninear pursuit-evasion game.

The rest of this paper is structured as follows. In Section \ref{sec:sec2}, the dynamical system models of an unpowered hypersonic pursuit vehicle and a maneuvering ground target are introduced to formulate a pursuit-evasion game. In Section \ref{sec:sec3}, an approximation of the saddle point strategy of the target is computed and subsequently, the pursuit-evasion game is reduced to an optimal control problem whose solution will serve as a reference pursuit trajectory. In Section \ref{sec:sec4}, the game dynamics are linearized with respect to the reference trajectory and an auxiliary two-player zero-sum LQDG is constructed. In Section \ref{sec:sim}, numerical simulations are presented. Lastly, in Section \ref{sec:concl}, some concluding remarks are provided.

\section{Two-Player Pursuit-Evasion Game} \label{sec:sec2}

In this section, the state space models that describe the motion of a hypersonic pursuit vehicle ($P$) and a maneuvering ground target ($T$) are introduced. Subsequently, the corresponding interception problem is formulated as a pursuit-evasion game.

\subsection{Problem setup and system models}
The game space, denoted by $\cX \subset \bR^2$, is defined as a vertical plane with the horizontal axis being the downrange and the vertical axis being the altitude, where $\cX := \{ (x,h) \in \bR^2 : h \geq 0 \}$. Let $\bm{\cX}_P := \cX \times \bR_{\geq 0} \times \bR$ be a state space and $\cU_P := [-\pi/18,\pi/18]$ denote a compact set of admissible inputs, then the non-affine, nonlinear equations of planar hypersonic motion of $P$ in $\cX$ are given by
\begin{align} \label{eq:kinematics_pursuer}
    \dot{\x}_P &= \f_P (\x_P,u_P), \quad \x_P(0) = \x_{P}^0,
\end{align}
where $\x_P \in \bm{\cX}_P$ (resp., $\x_P^0 \in \bm{\cX}_P$) is the state (resp., initial state), $u_P \in \cU_P$ is the control input, and $\f_P : \bm{\cX}_P \times \cU_P \rightarrow \bm{\cX}_P$ is the vector field of $P$. In particular, $\x_P$ includes
\begin{align}
    \dot x_P &= v_P \cos{\gamma_P},
    \\
    \dot h_P &= v_P \sin{\gamma_P},
    \\
    \dot v_P &= -\frac{D}{m} - g \sin{\gamma_P},
    \\
    \dot \gamma_P &= \frac{L}{mv_P}-\frac{g\cos{\gamma_P}}{v_P},
\end{align}
where $x_P$, $h_P$, $v_P$, and $\gamma_P$ denote the horizontal and vertical position, velocity, and flight path angle of $P$, respectively. Furthermore, $m$ is the vehicle mass, $g$ is gravitational acceleration, and $L$ and $D$ are the lift and drag forces which satisfy
\begin{align}
    L &= \frac{1}{2}\rho Sv_P^2 C_L,
    \\
    D &= \frac{1}{2}\rho Sv_P^2 C_D.
\end{align}
Here, $S$ is the vehicle reference area and $\rho$ is an exponential function that determines an approximation of the atmosphere density in terms of $P$'s vertical position, that is,
\begin{align}
    \rho & = \rho_0 e^{-h_P/H},
\end{align}
where $\rho_0$ is the surface air density and $H$ is the scale height. The lift and drag coefficients, $C_L$ and $C_D$, are known functions of the angle of attack $\alpha$:
\begin{align}
    C_L &= C_{L,1} \alpha,
    \\
    C_D &= C_{D,0} + C_{D,2} \alpha^2,
\end{align}
where $\alpha$ (angle of attack) is the control input of $P$, i.e., $u_P = \alpha$. Additionally, we denote the position of $P$ as $\bm{p}_P := (x_P,h_P) \in \cX$ (projection of $\x_P$ on $\cX$).

The target $T$ is a ground vehicle whose motion is constrained along the horizontal axis. In this paper, we assume the speed of $T$ to be much less than that of $P$ yet $T$ has better maneuverability. In particular, its motion is modelled by the so-called \textit{simple motion} kinematics \cite{isaacs1965differential} such that $T$ can directly control its velocity vector, that is,
\begin{align} \label{eq:kinematics_evader}
    \dot \x_T = \mbD u_T, \quad \x_T(0) = \x_T^0,
\end{align}
where $\x_T = (x_T,h_T) \in \cX$ (resp., $\x_T^0 = (x_T^0,h_T^0) \in \cX$) is the state (resp., initial state) and $u_T \in \cU_T := [-1,1]$ is the control input of $T$ with $x_T$ and $h_T$ denoting her horizontal and vertical position, respectively. Also, $\mbD = [v_T, 0]^\top$, where $v_T \in \bR_{> 0}$ denotes $T$'s maximum speed. Note that, since we assume the motion of $T$ is constrained along the horizontal axis, it holds that $h_T(t) = h_T^0 = 0, ~ \forall t \geq 0$.

\begin{figure}
    \centering
    \includegraphics[scale=0.4]{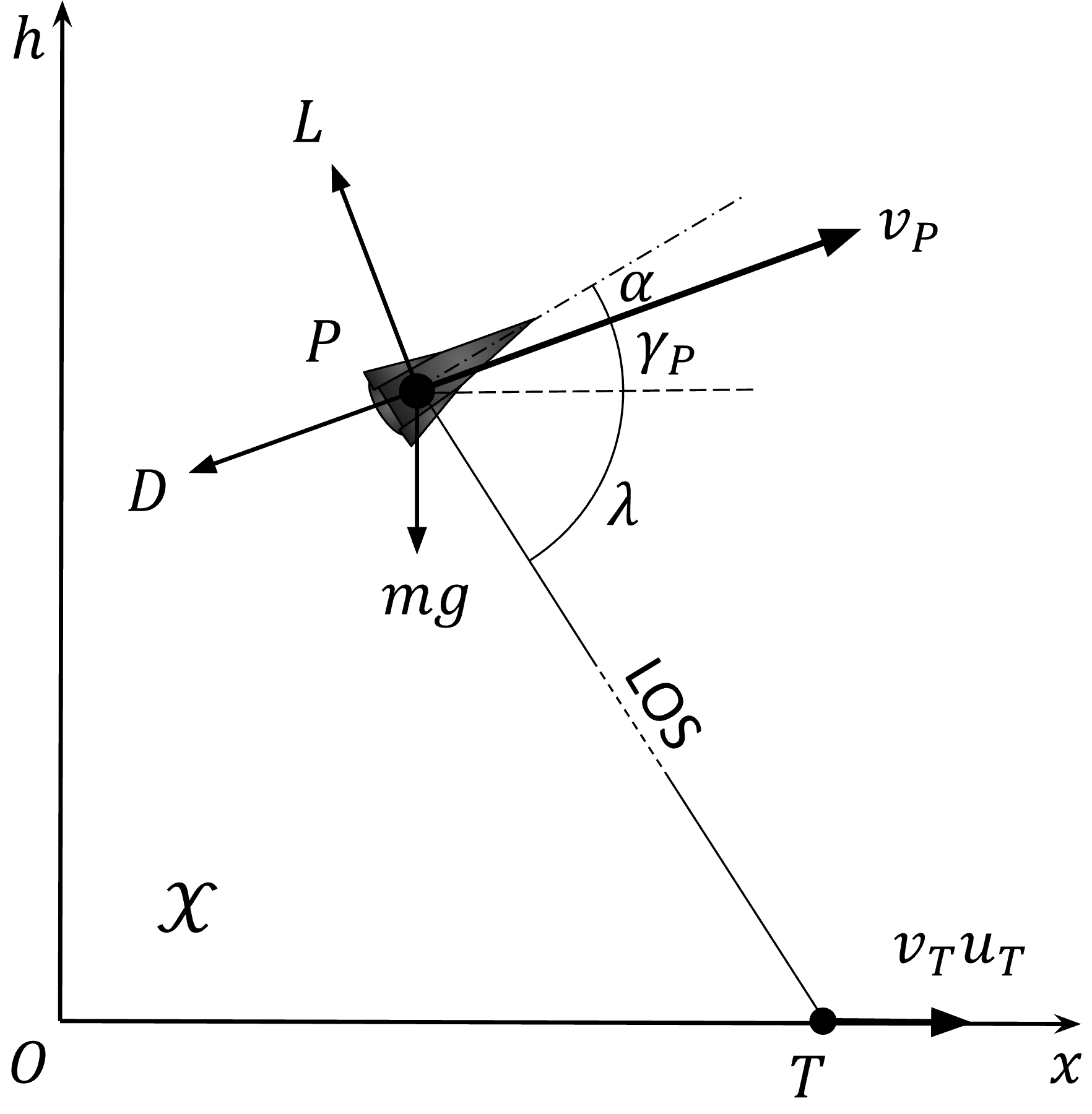}
    \caption{Planar pursuit-evasion engagement geometry}
    \label{fig:fig1}
\end{figure}

\subsection{The pursuit-evasion game}
In the target interception problem we will discuss, $P$'s goal is to intercept $T$ as soon as possible while $T$'s goal is to delay the interception. Hence, this interception problem can naturally be formulated as a pursuit-evasion game \cite{isaacs1965differential}:
\begin{problem}[Two-player pursuit-evasion game] \label{prob:prob1}
    \begin{equation}
        \begin{aligned}
            \min_{u_P(\cdot)} \max_{u_T(\cdot)}~~ & J = t_f
            \\
            \textrm{s.t.}~~ & \dot{\x}_P = \f_P (\x_P,u_P), \quad \x_P(0) = \x_{P}^0,
            \\
            & \dot{\x}_T = \mbD u_T, \qquad\qquad \x_T(0) = \x_{T}^0,
            \\
            & u_P \in \cU_P,~u_T \in \cU_T,
            \\
            &\Psi(\x_P(t_f),\x_T(t_f)) = 0, \nonumber
        \end{aligned}
    \end{equation}
\end{problem}
where $\Psi(\x_P,\x_T) := \| \x_T - \bm{p}_P \|_2 - \epsilon$ with $\epsilon$ denoting the capture radius. In words, given the system models \eqref{eq:kinematics_pursuer} and \eqref{eq:kinematics_evader} and the initial positions of $P$ and $T$, we are interested in finding the feedback strategies $u_P(\cdot)$ and $u_T(\cdot)$ such that, if these strategies are applied to \eqref{eq:kinematics_pursuer} and \eqref{eq:kinematics_evader}, $T$ will be intercepted by $P$ at time $t_f := \inf \{ t \in \bR_{\geq 0} : \Psi(\x_P(t_f),\x_T(t_f)) = 0 \}$ (time of capture). The planar geometry of pursuit-evasion between $P$ and $T$ is illustrated in Figure \ref{fig:fig1}, where all the variables therein have been defined except for $\lambda$ which denotes the angle of depression of $P$.

To ensure the feasibility of the game in Problem~\ref{prob:prob1}, we make a few assumptions:
\begin{assumption}
    (Perfect information game) Both $P$ and $T$ can observe their opponent's state at every instance of time.
\end{assumption}
\begin{assumption}
    (Game of kind) The initial states of both players are given such that capture is guaranteed under optimal play, i.e., $ \min_{u_P^\star(\cdot)} \max_{u_T^\star(\cdot)} J < \infty$.
\end{assumption}

The feedback strategies $u_P^\star(\cdot)$ and $u_T^\star(\cdot)$ constitute a (feedback) saddle point of the game must satisfy
\begin{equation}
    J(u_P^\star(\cdot),u_T(\cdot)) \leq J(u_P^\star(\cdot),u_T^\star(\cdot)) \leq J(u_P(\cdot),u_T^\star(\cdot)), \label{eq:sadpoint}
\end{equation}
which implies that if $P$ plays optimally while $T$ plays non-optimally, $P$ can take advantage of $T$'s non-optimal play and capture $T$ earlier than the saddle point time of capture. Conversely, if $P$ plays non-optimally while $T$ plays optimally, $T$ can delay (or even avoid) the termination of game. The game in Problem~\ref{prob:prob1} always possesses a saddle point that satisfies \eqref{eq:sadpoint} because the Hamiltonian of the game
\begin{align} \label{eq:hamiltonian}
    \cH = 1 + \bmlambda_P^\top \f_P + \bmlambda_T^\top \mbD u_T,
\end{align}
where $\bmlambda_P \in \bR^4$ and $\bmlambda_T \in \bR^2$ denote co-states, satisfies the Isaacs' condition~\cite{isaacs1965differential}:
\begin{align}
    \min_{u_P(\cdot)} \max_{u_T(\cdot)} \cH = \max_{u_T(\cdot)} \min_{u_P(\cdot)} \cH,
\end{align}
since \eqref{eq:hamiltonian} is separable in the control inputs $u_P$ and $u_T$. The Value of the game, which then is assured to exist, is given by
\begin{align}
    V = \min_{u_P(\cdot)} \max_{u_T(\cdot)} J = \max_{u_T(\cdot)} \min_{u_P(\cdot)} J.
\end{align}

\section{Open-Loop Pursuit-Evasion Trajectories} \label{sec:sec3}

In this section, we first derive the optimal strategy for $T$ in Problem~\ref{prob:prob1} by imposing a few additional assumptions on the dynamical behaviors of $P$ and $T$ in order to obtain an approximated value function. With this strategy we simplify the game in Problem~\ref{prob:prob1} into $P$'s one-sided optimal control problem and compute his optimal trajectory.

\subsection{Approximated saddle point strategy for optimal evasion}

First, we assume that $P$'s flight trajectory during the end game is a straight line, i.e., the flight path angle of $P$ aligns with its line of sight ($\gamma_P \approx \lambda$) and that the speed of $P$ linearly decreases with time. In consideration of these assumptions, the motion of $P$ can also be described by the simple motion kinematics (with decreasing maximum speed), and therefore the value function can be approximated as a linear function of the initial distance between the two players as follows:
\begin{align} \label{eq:value_approx}
    V := \min_{u_P^\star(\cdot)} \max_{u_T^\star(\cdot)} J = c\| \x_T - \bm{p}_P \|_2,
\end{align}
with $c = (\overline v_P - v_T)^{-1}$ where $\overline v_P$ denotes the average velocity of $P$ during the flight. The partial derivative of \eqref{eq:value_approx} with respect to $\x_T$ is given by
\begin{align} \label{eq:value_derivative}
    \frac{\partial V}{\partial \x_T} = \frac{c(\x_T - \overline \x_P)^\top}{\| \x_T - \overline \x_P \|_2}.
\end{align}
Substituting \eqref{eq:value_derivative} into the Hamilton-Jacobi-Isaacs (HJI) equation \cite{bacsar1998dynamic} and using the fact that the value function $V$ is time-invariant, one can derive
\begin{align}
    0 &= \min_{u_P \in \cU_P} \max_{u_T \in \cU_T} \left( 1 + \frac{\partial  V}{\partial \x_P} \f_P + \frac{\partial  V}{\partial \x_T} \mbD u_T \right) \nonumber
    \\
    &=  1 + \min_{u_P \in \cU_P} \frac{\partial  V}{\partial \x_P} \f_P + \max_{u_T \in \cU_T} \frac{2c v_T (x_T - x_P)^\top}{\| \x_T - \overline \x_P \|_2} u_T, \nonumber
\end{align}
from which we can conclude that the saddle point strategy for $T$ is given by
\begin{align} \label{eq:evader_feedback_strategy}
    &u_T^\star(\x_P(t),\x_T(t)) \nonumber
    \\
    &\quad =
    \begin{cases}
        \sgn(x_T(t)-x_P(t)), &\textrm{if}~x_T(t) \neq x_P(t),
        \\
        \textrm{undefined}, &\textrm{if}~x_T(t) = x_P(t).
    \end{cases}
\end{align}
In other words, the initial relative  position of $P$ with respect to $T$ in the $x$ coordinate determines the optimal heading direction of $T$. We can further assume that, excluding the case when $x_T = x_P$, no switching will occur during the game because, due to the assumption on the pursuit trajectory being a straight-line, the sign of relative $x$ position will never be switched. Thus, strategy \eqref{eq:evader_feedback_strategy} can also be considered an open-loop strategy for $T$.

\subsection{Optimal pursuit trajectory}

Despite $P$'s highly nonlinear equations of motion, given the fact that the optimal evasion trajectory of $T$ is a straight line with direction determined by the initial relative horizontal distance, the open-loop representation of the saddle point strategy of $P$ is no longer difficult to derive. Provided the initial positions of the two players, we first determine $T$'s (open-loop) strategy by $u_T^* = u_T^\star(\x_P^0,\x_T^0)$, which we call the Open-Loop Evasion Strategy (OLES). Since the evasion direction of $T$ is time-invariant and known \textit{a priori} to $P$, Problem~\ref{prob:prob1} can be reduced to a one-sided optimal control problem for $P$ as follows:
\begin{problem}[$P$'s optimal control problem] \label{prob:prob2}
    \begin{equation}
        \begin{aligned}
            \min_{u_P(\cdot)}~~ & J = t_f
            \\
            \textrm{s.t.}~~ & \dot{\x}_P = \f_P (\x_P,u_P), \quad \x_P(0) = \x_{P}^0,
            \\
            & \dot{\x}_T = \mbD u_T^*, \quad \x_T(0) = \x_{T}^0,
            \\
            & u_P \in \cU_P,
            \\
            &\Psi(\x_P(t_f),\x_T(t_f)) = 0. \nonumber
        \end{aligned}
    \end{equation}
\end{problem}
Problem \ref{prob:prob2} can be readily solved using any existing numerical optimal control technique (either direct or indirect). Details on the implementation of such techniques will be omitted herein due to space limitations, yet one can refer instead to standard references in the field (for instance, see~\cite{kelly2017introduction} and references therein). Such numerical techniques will yield the (nominal) minimum time of capture $t_f^*$ as well as the optimal state and input trajectories of $P$, denoted by $\x_P^*$ and $u_P^*$, respectively. We will also refer to $u_P^*$ as the Open-Loop Pursuit Strategy (OLPS). In addition, the state trajectories of $P$ and $T$ generated with the application of the OLPS and OLES, are together referred to as the Open-Loop Saddle Point State Trajectory (OLSPT).

\section{Auxiliary Zero-Sum LQDG} \label{sec:sec4}

The OLPS, albeit optimal in an open-loop sense, is in fact not effective against $T$ when the latter makes sub-optimal and / or decisions which are inconsistent with the OLES. This is because the OLPS cannot take into account the deviation in $T$'s state. In other words, as opposed to our discussion on \eqref{eq:sadpoint}, $P$ cannot take advantage of $T$'s sub-optimal play with the OLPS. To cope with the possible unexpected maneuvers of $T$, some auxiliary feedback input is needed such that the pursuit trajectory can be continuously reshaped in response to the state deviation of $T$.

Fortunately, in light of the fact that $P$ and $T$ have a large speed difference, we may further assume that any resulting pursuit-evasion trajectory caused by $T$'s evading maneuvers (whether optimal or not) will not significantly differ from the OLSPT. This particular assumption will allow us to linearize the dynamics of the players around the OLPTS and construct an auxiliary differential game in the neighborhood of this trajectory. Thereafter, we can derive an auxiliary feedback strategy for $P$ and combine it with the OLPS.

\subsection{Linear approximation of game dynamics}

To obtain an approximate linear state space model for the equations of motion of $P$ given in \eqref{eq:kinematics_pursuer} which is valid in the neighborhood of the OLSPT, we can take the first order Taylor series expansion of the vector field $\f_P$ and substitute the solution of Problem \ref{prob:prob2} (namely, $t_f^*$, $\x_P^*$, and $u_P^*$) into the resulting linearized equations. By doing so, we obtain the following linear time-varying (LTV) system model for $P$:
\begin{align} \label{eq:linearized_hypersonic_motion}
    \delta \dot \x_P &= \mbA(t) \delta \x_P + \mbB(t) \delta u_P,\quad \delta \x_P(0) = \delta \x_P^0,
\end{align}
where $\delta \x_P = \x_P - \x_P^*$, $\delta \x_P^0 = \x_P^0 - \x_P^{0 *}$, and $\delta u_P = u_P - u_P^*$ are the deviation of current state and input of $P$ from the open-loop ones, respectively. Moreover,
\begin{align}
    &\mbA(t) = \frac{\partial \f_P}{\partial \x_P} \bigg|_{\x_P^*(t),u_P^*(t)} = \nonumber
    \\
    &
    \begin{bmatrix}
        0 & 0 & \cos \gamma_P & - v_P \sin \gamma_P
        \\
        0 & 0 & \sin \gamma_P & v_P \cos \gamma_P
        \\
        0 & \frac{\kappa}{H} v_P^2 C_D & -2 \kappa v_P C_D & -g \cos \gamma_P
        \\
        0 & -\frac{\kappa}{H} v_P C_L & \kappa C_L + \frac{g \cos \gamma_P}{v_P^2} & \frac{g \sin \gamma_P}{v_P}
    \end{bmatrix},
    \\
    &\mbB(t) = \frac{\partial \f_P}{\partial u_P} \bigg|_{\x_P^*(t),u_P^*(t)} = \begin{bmatrix}
        0 \\ 0 \\ -2\kappa C_{D,2} v_P^2 \alpha \\ \kappa C_{L,1} v_P
    \end{bmatrix},
\end{align}
where $\kappa := S \rho/(2m)$. Note that the matrices $\mbA(t)$ and $\mbB(t)$ are only defined over the finite time interval $[0,t_f^*]$.

Similarly, $T$'s motion near her optimal evasion trajectory can be derived (in an almost trivial way) as
\begin{align} \label{eq:linearized_target_motion}
    \delta \dot \x_T &= \mbD \delta u_T,\quad \delta \x_T(0) = \delta \x_T^0,
\end{align}
which is in the same form as \eqref{eq:kinematics_evader}.

\subsection{Auxiliary differential game}

Now, given the linear system models \eqref{eq:linearized_hypersonic_motion} and \eqref{eq:linearized_target_motion}, an auxiliary differential game can be constructed around the OLSPT. The corresponding game dynamics are written as
\begin{align} \label{eq:linearized_game_dynamics}
    \dot{\mbx} &= \bm{\cA}(t) \mbx + \bm{\cB}(t) \nu_P + \bm{\cD} \nu_T, \quad \mbx(0) = \mbx_0,
\end{align}
where $\mbx = (\delta \x_T, \delta \x_P)$ is the joint state deviation, $\mbx_0 = (\delta \x_T^0, \delta \x_P^0)$ is the initial joint state deviation, $\nu_P = \delta u_P$ and $\nu_T = \delta u_T$, and
\begin{align}
    \bm{\cA}(t) &=
    \left[
    \begin{array}{c|c}
    \bm{0}_{2 \times 2}
    &
    \bm{0}_{2 \times 4}
    \\
    \hline
    \bm{0}_{4 \times 2}
    &
    \mbA(t)
    \end{array}
    \right],
    \\
    \bm{\cB}(t) &=
    \begin{bmatrix}
        \bm{0}_{2 \times 1}
        \\
        \mbB(t)
    \end{bmatrix},
    \\
    \bm{\cD} &=
    \begin{bmatrix}
        \mbD \\ \bm{0}_{4 \times 1}
    \end{bmatrix}.
\end{align}
Note again that these matrices are also defined over the finite interval $[0,t_f^*]$ only. Furthermore, strategies $\nu_P$ and $\nu_T$ are not bounded in this formulation. With this LTV game dynamics, we construct a two-player zero-sum LQDG:
\begin{problem}[Auxiliary two-player zero-sum LQDG] \label{prob:auxLQDG}
    \begin{equation}
        \begin{aligned}
            \min_{\nu_P(\cdot)}\max_{\nu_T(\cdot)}~~ & \cJ = w_1 \left[ \left( \delta x_T - \delta x_P \right)^2 + w_2 \left( \delta h_T - \delta h_P \right)^2 \right]
            \\
            &\qquad\qquad\qquad + \int_{0}^{t_f^*}\left[ \nu_P^2 - w_3 \nu_T^2(t) \right] \ud t
            \\
            \textrm{s.t.}~~ & \dot \mbx = \bm{\cA}(t) \mbx + \bm{\cB}(t) \nu_P + \bm{\cD} \nu_T, ~~ \mbx(0) = \mbx_0, \nonumber
        \end{aligned}
    \end{equation}
\end{problem}
where the auxiliary payoff functional $\cJ$ is the weighted combination of the soft terminal constraint on miss distance and the accumulated control inputs. The weight coefficients $w_1$, $w_2$, and $w_3$ are positive constants which can be tuned via trial and error. The fixed final time of the game corresponds to the nominal time of capture $t_f^*$ which we have already found in Problem~\ref{prob:prob2}. Let us additionally define a matrix
\begin{align}
    \mbQ &= w_1
    \left[
    \begin{array}{c|c}
    \begin{matrix}
        1 & 0 & -1 & 0
        \\
        0 & w_2 & 0 & -w_2
        \\
        -1 & 0 & 1 & 0
        \\
        0 & -w_2 & 0 & w_2
    \end{matrix}
    &
    \bm{0}_{4 \times 2}
    \\
    \hline
    \bm{0}_{2 \times 4}
    &
    \bm{0}_{2 \times 2}
    \end{array}
    \right],
\end{align}
then the auxiliary payoff functional can equivalently be written as
\begin{align}
    \cJ = \mbx^\top(t_f^*) \mbQ \mbx(t_f^*) + \int_{0}^{t_f^*}\left[ \nu_P^2 - w_3 \nu_T^2(t) \right] \ud t.
\end{align}
The feedback saddle point solution of Problem~\ref{prob:auxLQDG}, namely $\nu_P^\star$ and $\nu_T^\star$ which satisfy a similar inequality as \eqref{eq:sadpoint}:
\begin{align}
    \cJ(\nu_P^\star(\cdot),\nu_T(\cdot)) \leq \cJ(\nu_P^\star(\cdot),\nu_T^\star(\cdot)) \leq \cJ(\nu_P(\cdot),\nu_T^\star(\cdot)) \nonumber
\end{align}
can readily be derived as \cite{li2011defending}
\begin{align}
    \nu_P^\star(\mbx(t),t) &= - \bm{\cB}^\top(t) \mbP(t) \mbx(t), \label{eq:nu_P}
    \\
    \nu_T^\star(\mbx(t),t) &= w_3^{-1} \bm{\cD}^\top \mbP(t) \mbx(t), \label{eq:nu_T}
\end{align}
where $\mbP$ corresponds to the solution of the following Matrix Riccati Differential Equation (MRDE):
\begin{align} \label{eq:mrde}
    -\dot\mbP = \bm{\cA}^\top(t) \mbP + \mbP \bm{\cA}^\top(t) - \mbP \bm{\cS}(t) \mbP, ~~ \mbP(t_f^*) = \mbQ,
\end{align}
with
\begin{align}
    \bm{\cS}(t) := \bm{\cB}(t) \bm{\cB}^\top(t) - w_3^{-1} \bm{\cD} \bm{\cD}^\top.
\end{align}
By solving the MRDE backward in time one can compute the feedback gains in \eqref{eq:nu_P} and \eqref{eq:nu_T}. Finally, to ensure that the aggregate feedback strategies meet the input constraints defined in Section 1, we define a function $\phi_i : \bR \rightarrow \cU_i$ for $i \in \{P,T\}$ such that
\begin{align}
    \phi_i(u) :=
    \begin{cases}
        u, &\textrm{if}~u \in \cU_i,
        \\
        \sup~ \cU_i, &\textrm{if}~u \geq \sup~ \cU_i,
        \\
        \inf~ \cU_i, &\textrm{if}~u \leq \inf~ \cU_i,
    \end{cases}
\end{align}
where $u \in \bR$. Hence, the (bounded) aggregate feedback strategies are given by
\begin{align}
    \widetilde{u}_P^\star(\mbx(t),t) &= \phi_P (u_P^*(t) + \nu_P^\star(\mbx(t),t)), \label{eq:final_law_P}
    \\
    \widetilde{u}_T^\star(\mbx(t),t) &= \phi_T ( u_T^*(t) + \nu_T^\star(\mbx(t),t)). \label{eq:final_law_T}
\end{align}

\begin{table}
\renewcommand{\arraystretch}{1.3}
\caption{\bf Hypersonic vehicle parameters}
\label{tab:tab1}
\centering
\begin{tabular}{|c|c|c|}
\hline
\bfseries Parameter & \bfseries Symbol & \bfseries Value (units) \\
\hline\hline
Reference area & $S$ & 0.2919 $m^2$ \\
Mass & $m$ & 340.1943 $kg$ \\
Lift coefficient & $C_{L,1}$ & 1.5658
\\
Drag coefficients & $C_{D,0}$, $C_{D,2}$ & 0.0612, 1.6537
\\
\hline
\end{tabular}
\end{table}

\begin{table}
\renewcommand{\arraystretch}{1.3}
\caption{\bf Initial state variables of players}
\label{tab:tab2}
\centering
\begin{tabular}{|c|c|}
\hline
\bfseries Variable & \bfseries Initial value (units) \\
\hline\hline
$x_P^0$ & $-$50,000 $m$ \\
$h_P^0$ & 20,000 $m$ \\
$v_P^0$ & 4,000 $m$ \\
$\gamma_P^0$ & $-$0.4 $rad$ \\
$x_T^0$ & 0 $m$ \\
$h_T^0$ & 0 $m$ \\
\hline
\end{tabular}
\end{table}

\begin{table}
\renewcommand{\arraystretch}{1.3}
\caption{\bf Different evasion strategies and miss distance}
\label{tab:tab3}
\centering
\begin{tabular}{|c|c|}
\hline
\bfseries Evasion strategy & \bfseries Miss distance (units) \\
\hline\hline
$E_1$ & 13.1660 $m$ \\
$E_2$ & 16.8271 $m$ \\
$E_3$ & 3.1349 $m$ \\
\hline
\end{tabular}
\end{table}

\section{Numerical Simulations} \label{sec:sim}

In this section, we present numerical simulation results of the proposed game-theoretic terminal guidance law. The vehicle parameters of $P$ are specified in Table \ref{tab:tab1}, whereas the initial conditions of $P$ and $T$ are listed in Table \ref{tab:tab2}. Other parameters that are necessary for the simulations are selected as: surface air density $\rho_0= 1.2 ~kg/m^3$, scale height $H = 7,500 ~m$, and maximum target speed $v_T = 20 ~ m/s$. The angle of attack, $\alpha$, takes values in the compact set $\mathcal{U}_P:= [ -\pi/\mathrm{18}, \pi / \mathrm{18} ]$. The capture radius $\epsilon$ is assumed to be zero, which means that the game terminates if the positions of $P$ and $T$ coincide.

\begin{figure}
    \centering
    \includegraphics[scale=0.37]{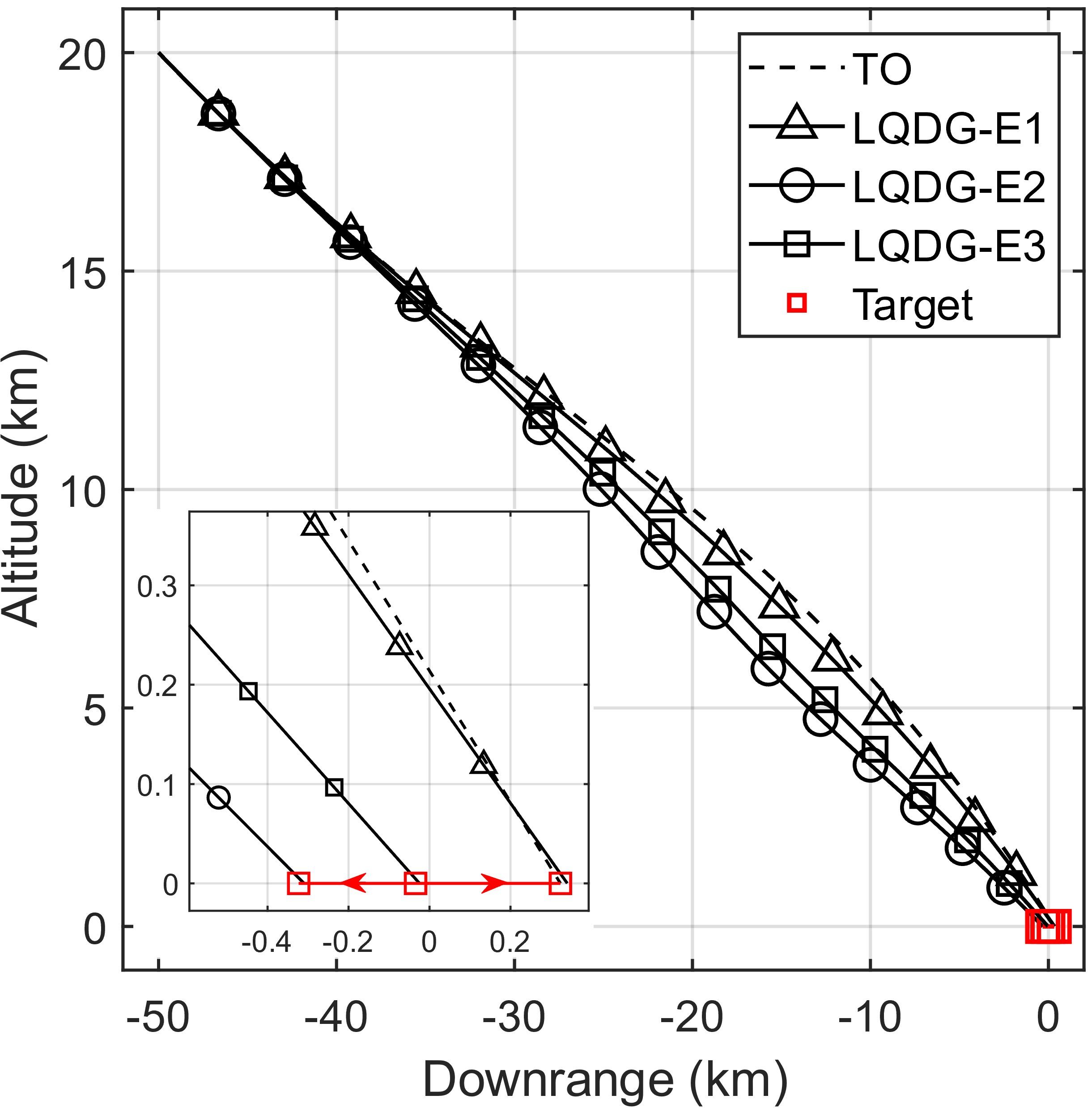}
    \caption{Pursuit-evasion trajectories}
    \label{fig:trajectories}
\end{figure}

\begin{figure}
    \centering
    \includegraphics[scale=0.25]{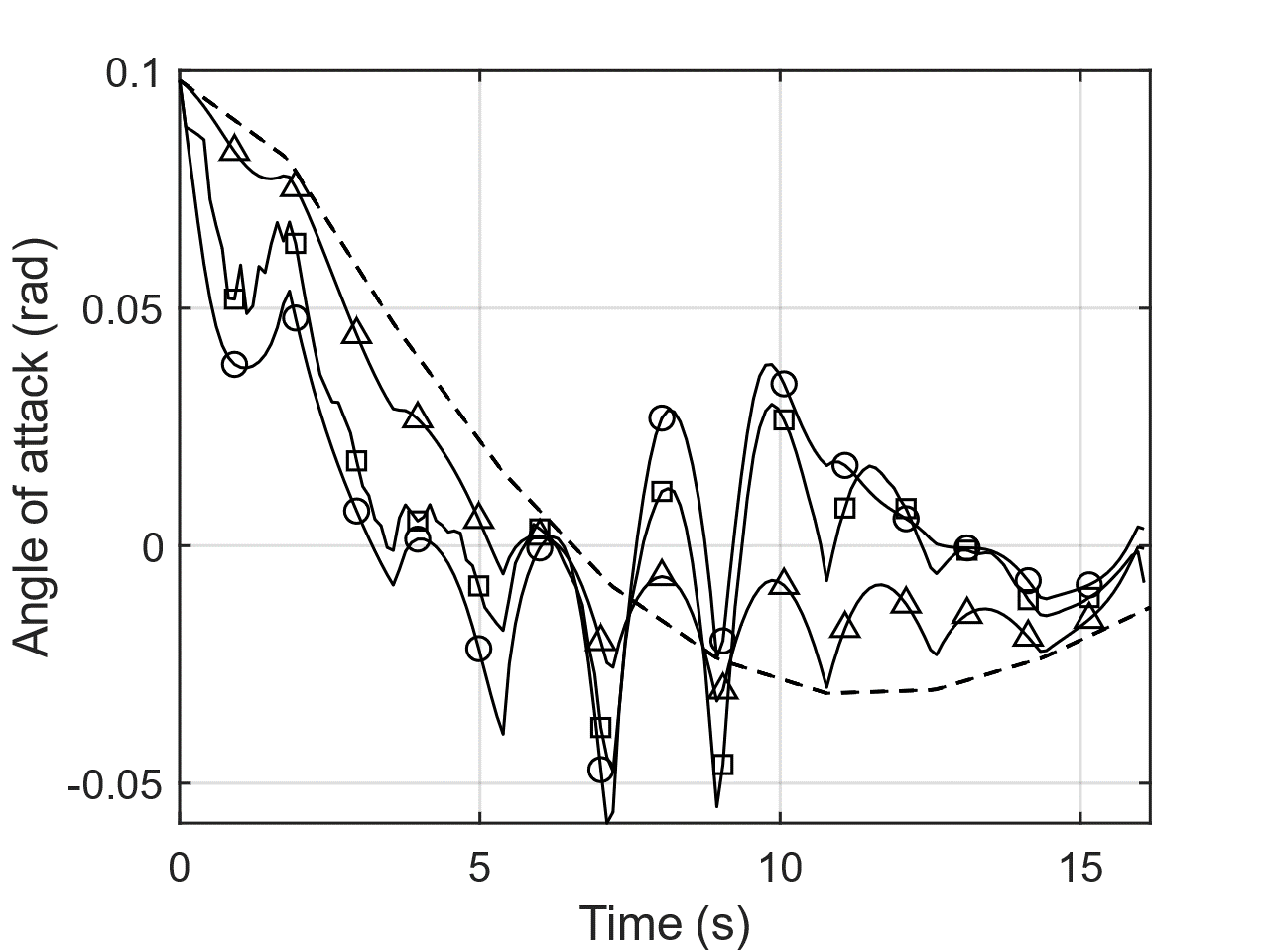}
    \caption{Input trajectories (angle of attack)}
    \label{fig:inputs}
\end{figure}

We first solve Problem \ref{prob:prob2} by utilizing a numerical optimal control technique (for instance, direct collocation \cite{kelly2017introduction}) to obtain the OLPS. Note that one can use any numerical method here (either direct or indirect), and in fact the OLPS and the corresponding reference trajectory do not have to be optimal (i.e., they may be sub-optimal) in order to apply the aggregate guidance laws proposed in Section~\ref{sec:sec4}; however, the closer this reference trajectory is to the optimal one, the smaller $t_f^*$ one can obtain. From Table \ref{tab:tab2} we see that $T$ is initially located to the right of $P$ (i.e., $x_P^0 < x_T^0$), which implies that the OLES, according to \eqref{eq:evader_feedback_strategy}, is given by $u_T^* = 1$. In the following simulation results, we refer to $P$'s optimal trajectory, $\x_P^*$, as the Time Optimal (TO) trajectory, which is illustrated as a dash line in Figure \ref{fig:trajectories}. In addition, the nominal time of capture is obtained from the numerical solver as $t_f^* = 16.1625 (s)$.

\begin{figure*}
    \centering
    \includegraphics[scale=0.22]{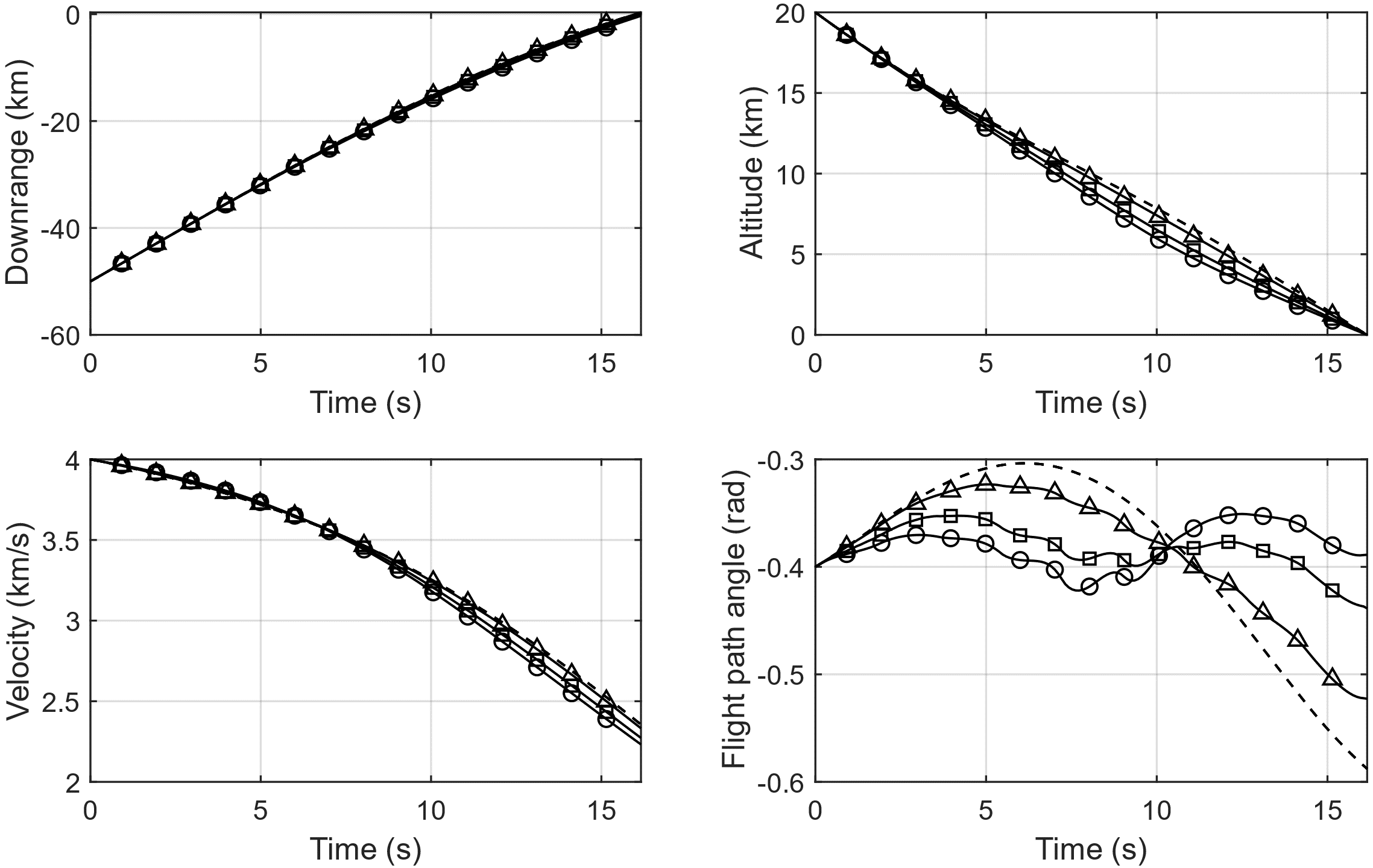}
    \caption{State trajectories of the hypersonic vehicle}
    \label{fig:states}
\end{figure*}

Given the TO trajectory and the nominal time of capture, we can compute the time-varying matrices $\mbA(t)$ and $\mbB(t)$ as well as $\bm{\cA}(t)$ and $\bm{\cB}(t)$ prior to the commencement of game, all of which are defined over the finite time horizon $[0,t_f^*]$. Given these matrices, the MRDE \eqref{eq:mrde} can be solved numerically (backward as said) to compute the feedback gains in \eqref{eq:nu_P} and \eqref{eq:nu_T}, also prior to the commencement of game. Proper selection of the weighted coefficients, $w_1$, $w_2$, and $w_3$, is the most critical process of this guidance algorithm, which must be done via trial and error. The tuning procedures we have adopted are as follows; first, we find a sufficiently smaller value of $w_1$ from its current value such that the auxiliary control input $\nu_P$ does not override the OLPS $u_P^*$, in which case $P$ can easily stall. Once this is done, the final position of $P$ may be located at the desired $x$ position but not at the proper $h$ position (non-zero altitude). For this reason, we next increase the value of $w_2$ such that $\delta h_P \rightarrow 0$. Our final choice for the weighted coefficients are as follows: $w_1 = 3 \cdot 10^{-5}$, $w_2 = 10^3$, and $w_3 = 10^3$.

In order to investigate the performance of our guidance law, namely \eqref{eq:final_law_P}, against unpredictable evasion of $T$, we apply it against $T$ who employs one of the following three evasion strategies: The first strategy, named $E1$, is the optimal evasion, i.e., $u_T(t) = u_T^*$ for all $t \in [0, t_f^*]$. In this case, we expect that $P$ will simply follow (or track) its reference trajectory since the deviation of $T$'s horizontal position ($\delta x_T$) will always be zero. The second strategy, $E2$, is evasion in the opposite direction of $E1$, i.e., $u_T(t) = -u_T^*$ for all $t \in [0, t_f^*]$. From the game-theoretic point of view, this strategy would be considered the worst strategy for $T$ since it would yield the smallest time of capture, only if $P$ knew its feedback saddle point strategy. Since $P$ employs, however, the proposed guidance law which is not necessarily optimal in Problem \ref{prob:prob1}, $E2$ can be used to test the robustness of the latter guidance law. Lastly, the third strategy, named $E3$, is random evasion which is meant to confuse $P$ by changing the evasion direction randomly; in particular, $T$ periodically chooses her control input to be a random sample from the (discrete) uniform distribution over the set $\{-1,0,1\}$.

In Figure~\ref{fig:trajectories}, the trajectories of $T$ corresponding to $E1$, $E2$, and $E3$ are shown as red curves, whereas the trajectories of $P$ in response to these strategies are shown as black curves with different shapes of markers. The miss distances are listed in Table \ref{tab:tab3}, all of which turn out to be reasonably small. Since our terminal guidance law is prescribed with a fixed final time, in all three cases $P$ reaches his final position at $t_f^*$. It is interesting to observe that $P$'s trajectory against $E1$ is, as opposed to our earlier expectation, is notably different from the TO trajectory. This is presumably because $P$ employs the OLPS, originally a smooth function obtained from the numerical solver, as a piece-wise continuous function in the numerical simulations. Despite the numerical error that comes from such discretization, however, it is shown that $P$ is able to intercept $T$ eventually with a reasonably small miss distance. Finally, the input and state trajectories of $P$ are shown in Figures~\ref{fig:inputs} and \ref{fig:states}, respectively.

\section{Conclusions} \label{sec:concl}
In this paper, we have proposed a terminal feedback guidance law that is based on a linearized approximation of the nonlinear hypersonic dynamics and the saddle point solution to a two-player zero-sum linear quadratic differential game. In the proposed guidance algorithm, given that the target's optimal evasion is confined in one axis and under a few simplifying assumptions, an open-loop reference pursuit trajectory have been computed by solving an optimal hypersonic trajectory optimization problem. The latter trajectory has then been used to linearize the nonlinear game dynamics and construct an auxiliary linear quadratic differential game whose saddle point solution is combined with the open-loop input. We have also presented numerical simulation results to verify that, as long as the weighted coefficients are properly selected via trial and error, our proposed guidance law performs well and guarantees the hypersonic pursuit vehicle to intercept the maneuvering (evading) target in a finite time with reasonably small miss distances. One of the limitations of the proposed scheme is that it is only applicable to a scenario in which the target's motion is constrained along a line. It is also required that the target's position must be known to the hypersonic pursuit vehicle at every time instant in order to compute the feedback control input derived from the auxiliary differential game along a reference pursuit trajectory. Possible future directions of this work include consideration of scenarios in which the target may evade on a plane and the information given to the hypersonic pursuit vehicle on the target's position is stochastic rather than deterministic.








\bibliographystyle{IEEEtran}
\bibliography{pegref}

\thebiography

\begin{biographywithpic}
{Yoonjae Lee}{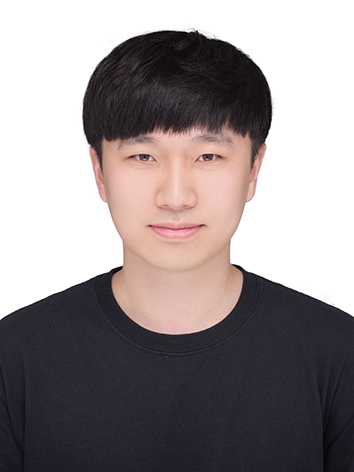}
received the B.S. degree in aerospace engineering from the University of California, San Diego, CA, USA in 2020. He is currently enrolled as a graduate student studying aerospace engineering in the Department of Aerospace Engineering and Engineering Mechanics at the University of Texas at Austin. His research is mainly focused on game theory, multi-agent systems, and pursuit-evasion games.
\end{biographywithpic}

\begin{biographywithpic}
{Efstathios Bakolas (Member, IEEE)}{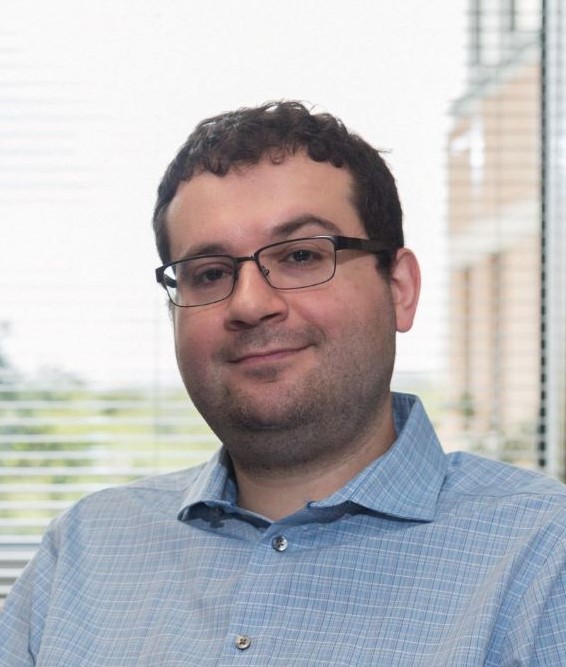}
received the Diploma degree in mechanical engineering with highest honors from the 
National Technical University of Athens, Athens, Greece, in 2004 and the M.S. and Ph.D. degrees in aerospace engineering from the Georgia Institute of Technology,
Atlanta, Atlanta, GA, USA, in 2007 and 2011, respectively. He is currently an Associate Professor with the Department of Aerospace Engineering and
Engineering Mechanics, The University of Texas at Austin, Austin, TX, USA. His research is mainly focused on stochastic
optimal control theory, optimal decision-making, differential and dynamic games, control of uncertain systems, data-driven modeling and control of 
nonlinear systems, and control of aerospace systems.
\end{biographywithpic}

\begin{biographywithpic}
{Maruthi R. Akella}{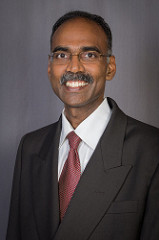}
is a tenured faculty member with the Department of Aerospace Engineering and Engineering Mechanics at The University of Texas at Austin (UT Austin) where he holds the Ashley H. Priddy Centennial Professorship in Engineering. He is the founding director for the Center for Autonomous Air Mobility and the faculty lead for the Control, Autonomy, and Robotics area at UT Austin. His research program encompasses control theoretic investigations of nonlinear and coordinated systems, vision-based sensing for state estimation, and development of integrated human and autonomous multivehicle systems. He is Editor-in-Chief for the Journal of the Astronautical Sciences and previously served as the Technical Editor (Space Systems) for the IEEE Transactions on Aerospace and Electronic Systems. He is a Fellow of the IEEE and AAS.
\end{biographywithpic}

\end{document}